\documentclass[aps,prd,twocolumn,preprint,tightenlines,amssymb,amsmath,superscriptaddress,showpacs,byrevtex, 10pt]{revtex4}
\usepackage{graphicx} % Include figure files
\usepackage{dcolumn}  % Align table columns on decimal point
\usepackage{multirow}
\usepackage{color}

\graphicspath{{ps}}

\newcommand{\BR}{$\mathcal{B}$}
\newcommand{\brIs}[2]{$\mathcal{B}(#1) = #2 $}
\newcommand{\kshort}{$K_S^0$}
\newcommand{\kstar}{$\bar{K}^*(892)^0$}
\newcommand{\dsub}[1]{$D_s^{#1}$}
\newcommand{\dsst}[1]{$D_s^{*#1}$}
\newcommand{\phiMode}{$\phi\pi$ mode}
\newcommand{\kstarMode}{$\bar{K}^*(892)^0K$ mode}
\newcommand{\kshortMode}{$K_S^0K$ mode}
\newcommand{\dsstPi}{$B^0 \to D_s^{*+}\pi^-$}
\newcommand{\dsstK}{$B^0 \to D_s^{*-}K^+$}
\newcommand{\resDsstPi}{\brIs{B^0 \to D_s^{*+} \pi^-}{(1.75 \pm 0.34 \textrm{ (stat)} \pm 0.17 \textrm{ (syst)} \pm 0.11 \textrm{ (\BR)}) \times 10^{-5}}}
\newcommand{\resDsstK}{\brIs{B^0 \to D_s^{*-} K^+}{(2.02 \pm 0.33 \textrm{ (stat)} \pm 0.18 \textrm{ (syst)} \pm 0.13 \textrm{ (\BR)}) \times 10^{-5}}}

\begin{document}

\preprint{\vbox{ \hbox{ } 
    \hbox{Belle Preprint {\it 2009-25}}
    \hbox{KEK Preprint {\it 2009-33}} 
    % \hbox{hep-ex nnnn} 
}}

\title{ \quad\\[1.0cm] Measurement of the branching fractions for
\dsstPi~and \dsstK~decays}

%% author list
%%% Paper:    B0 -> D_s* h
%%% Journal:  Physical Review D (Rapid Communication)
%%% Contacts: N.J. Joshi (njoshi@bpost.kek.jp)
%%%           K. Trabelsi (karim.trabelsi@kek.jp)
%%%           T. Aziz (aziz@tifr.res.in)
%%% Non-responding authors or those who said NO are commented out.
%%% ====================================================================
%%% Click the RELOAD button on your web browser to see the updated file.
%%% ====================================================================
%%% Use \input{author} to insert this material into your latex file.
%%%%% Force institutions to appear in alphabetical order when typeset.
\affiliation{Budker Institute of Nuclear Physics, Novosibirsk}
\affiliation{Faculty of Mathematics and Physics, Charles University, Prague}
%%%\affiliation{Chiba University, Chiba}
\affiliation{University of Cincinnati, Cincinnati, Ohio 45221}
\affiliation{Department of Physics, Fu Jen Catholic University, Taipei}
%%%\affiliation{Justus-Liebig-Universit\"at Gie\ss{}en, Gie\ss{}en}
\affiliation{The Graduate University for Advanced Studies, Hayama}
%%%\affiliation{Gyeongsang National University, Chinju}
\affiliation{Hanyang University, Seoul}
\affiliation{University of Hawaii, Honolulu, Hawaii 96822}
\affiliation{High Energy Accelerator Research Organization (KEK), Tsukuba}
%%%\affiliation{Hiroshima Institute of Technology, Hiroshima}
%%%\affiliation{University of Illinois at Urbana-Champaign, Urbana, Illinois 61801}
\affiliation{Institute of High Energy Physics, Chinese Academy of Sciences, Beijing}
\affiliation{Institute of High Energy Physics, Vienna}
\affiliation{Institute of High Energy Physics, Protvino}
%%%\affiliation{Institute of Mathematical Sciences, Chennai}
%%%\affiliation{INFN - Sezione di Torino, Torino}
\affiliation{Institute for Theoretical and Experimental Physics, Moscow}
\affiliation{J. Stefan Institute, Ljubljana}
\affiliation{Kanagawa University, Yokohama}
\affiliation{Institut f\"ur Experimentelle Kernphysik, Karlsruhe Institut f\"ur Technologie, Karlsruhe}
\affiliation{Korea University, Seoul}
%%%\affiliation{Kyoto University, Kyoto}
\affiliation{Kyungpook National University, Taegu}
\affiliation{\'Ecole Polytechnique F\'ed\'erale de Lausanne (EPFL), Lausanne}
\affiliation{Faculty of Mathematics and Physics, University of Ljubljana, Ljubljana}
\affiliation{University of Maribor, Maribor}
\affiliation{Max-Planck-Institut f\"ur Physik, M\"unchen}
\affiliation{University of Melbourne, School of Physics, Victoria 3010}
\affiliation{Nagoya University, Nagoya}
%%%\affiliation{Nara University of Education, Nara}
\affiliation{Nara Women's University, Nara}
\affiliation{National Central University, Chung-li}
\affiliation{National United University, Miao Li}
\affiliation{Department of Physics, National Taiwan University, Taipei}
\affiliation{H. Niewodniczanski Institute of Nuclear Physics, Krakow}
\affiliation{Nippon Dental University, Niigata}
\affiliation{Niigata University, Niigata}
\affiliation{University of Nova Gorica, Nova Gorica}
\affiliation{Novosibirsk State University, Novosibirsk}
\affiliation{Osaka City University, Osaka}
%%%\affiliation{Osaka University, Osaka}
\affiliation{Panjab University, Chandigarh}
%%%\affiliation{Peking University, Beijing}
%%%\affiliation{Princeton University, Princeton, New Jersey 08544}
%%%\affiliation{RIKEN BNL Research Center, Upton, New York 11973}
%%%\affiliation{Saga University, Saga}
\affiliation{University of Science and Technology of China, Hefei}
\affiliation{Seoul National University, Seoul}
%%%\affiliation{Shinshu University, Nagano}
\affiliation{Sungkyunkwan University, Suwon}
\affiliation{School of Physics, University of Sydney, NSW 2006}
\affiliation{Tata Institute of Fundamental Research, Mumbai}
\affiliation{Excellence Cluster Universe, Technische Universit\"at M\"unchen, Garching}
%%%\affiliation{Toho University, Funabashi}
\affiliation{Tohoku Gakuin University, Tagajo}
\affiliation{Tohoku University, Sendai}
\affiliation{Department of Physics, University of Tokyo, Tokyo}
%%%\affiliation{Tokyo Institute of Technology, Tokyo}
\affiliation{Tokyo Metropolitan University, Tokyo}
\affiliation{Tokyo University of Agriculture and Technology, Tokyo}
%%%\affiliation{Toyama National College of Maritime Technology, Toyama}
\affiliation{IPNAS, Virginia Polytechnic Institute and State University, Blacksburg, Virginia 24061}
\affiliation{Yonsei University, Seoul}

  \author{N.~J.~Joshi}\affiliation{Tata Institute of Fundamental Research, Mumbai} % Tata
  \author{T.~Aziz}\affiliation{Tata Institute of Fundamental Research, Mumbai} % Tata
  \author{K.~Trabelsi}\affiliation{High Energy Accelerator Research Organization (KEK), Tsukuba} % KEK
  \author{I.~Adachi}\affiliation{High Energy Accelerator Research Organization (KEK), Tsukuba} % KEK
  \author{H.~Aihara}\affiliation{Department of Physics, University of Tokyo, Tokyo} % Tokyo
  \author{K.~Arinstein}\affiliation{Budker Institute of Nuclear Physics, Novosibirsk}\affiliation{Novosibirsk State University, Novosibirsk} % BINP
% \author{T.~Aso}\affiliation{Toyama National College of Maritime Technology, Toyama} % Toyama
  \author{V.~Aulchenko}\affiliation{Budker Institute of Nuclear Physics, Novosibirsk}\affiliation{Novosibirsk State University, Novosibirsk} % BINP
  \author{T.~Aushev}\affiliation{\'Ecole Polytechnique F\'ed\'erale de Lausanne (EPFL), Lausanne}\affiliation{Institute for Theoretical and Experimental Physics, Moscow} % ITEP
  % \author{S.~Bahinipati}\affiliation{University of Cincinnati, Cincinnati, Ohio 45221} % Cincinnati
  \author{A.~M.~Bakich}\affiliation{School of Physics, University of Sydney, NSW 2006} % Sydney
  \author{V.~Balagura}\affiliation{Institute for Theoretical and Experimental Physics, Moscow} % ITEP
% \author{Y.~Ban}\affiliation{Peking University, Beijing} % Peking
% ...
% BINP
\collaboration{The Belle Collaboration}
%% end author list

\begin{abstract}
We present a study of \dsstPi and \dsstK decays based on a sample of
 $657 \times 10^6$ $B\bar{B}$ events collected with the Belle detector at
 the KEKB asymmetric-energy $e^+e^-$ collider. We measure the
 branching fractions to be \resDsstPi~and \resDsstK, with
 significances of 6.1 and 8.0 standard deviations, respectively. The
 first uncertainty is statistical, the second is due to the
 experimental systematics, and the third is from uncertainties in the
 \dsub{+}~decay branching fractions. From our measurements, we obtain
 the most precise determination so far of $R_{D^*\pi}$, where
 $R_{D^*\pi}$ is the ratio between amplitudes of the doubly
 Cabibbo-suppressed decay $B^0 \to D^{*+}\pi^-$ and the Cabibbo
 favored $B^0 \to D^{*-}\pi^+$ decay.

\end{abstract}

\pacs{11.30.Er, 12.15.Hh, 13.25.Hw}

\maketitle

%%%% >>>> keep the final version single-spaced
\tighten

{\renewcommand{\thefootnote}{\fnsymbol{footnote}}}
\setcounter{footnote}{0}

\begin{figure}[h!]
\includegraphics[width=0.22\textwidth]{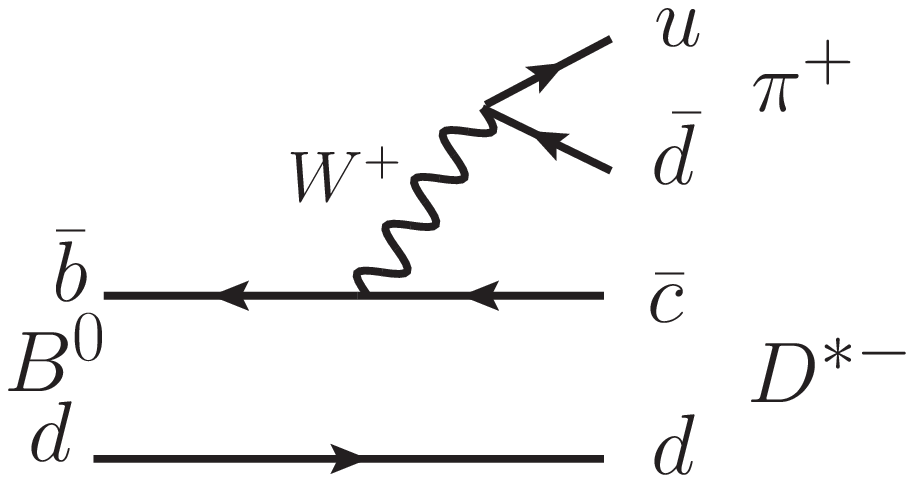}
\includegraphics[width=0.25\textwidth]{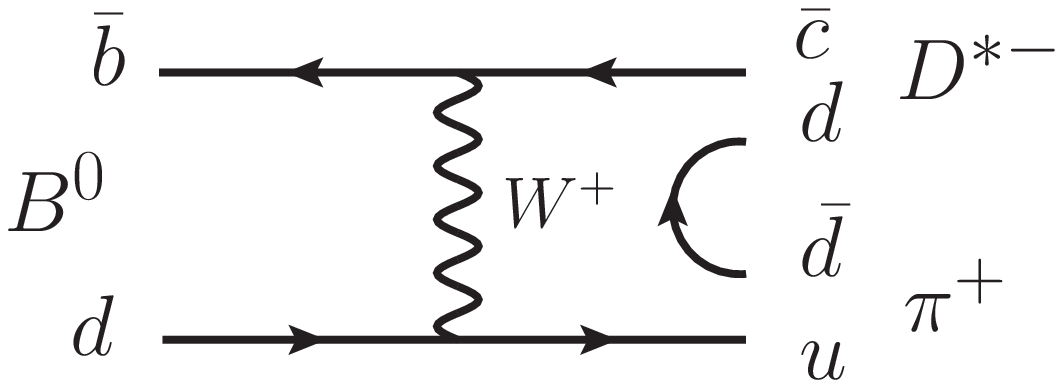}\\
\textrm{\hskip -1cm (a) \hskip 4cm (d)}

\includegraphics[width=0.22\textwidth]{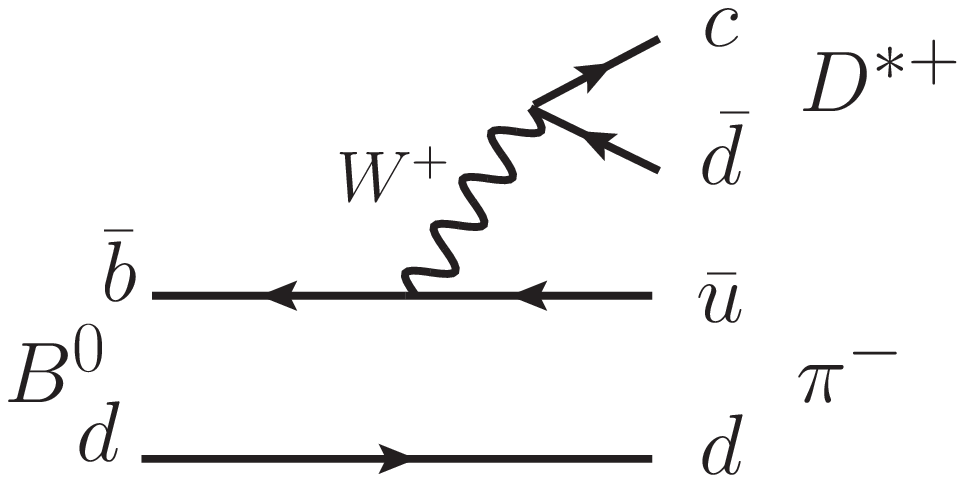}
\includegraphics[width=0.25\textwidth]{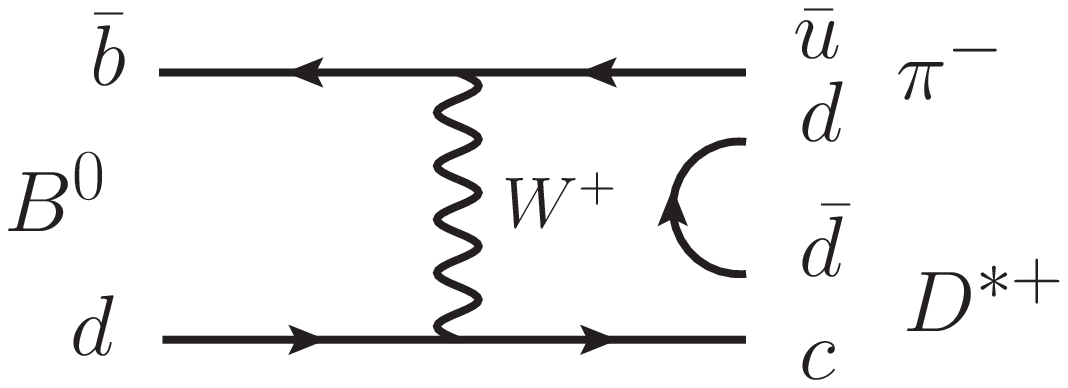}\\
\textrm{\hskip -1cm (b) \hskip 4cm (e)}

\includegraphics[width=0.22\textwidth]{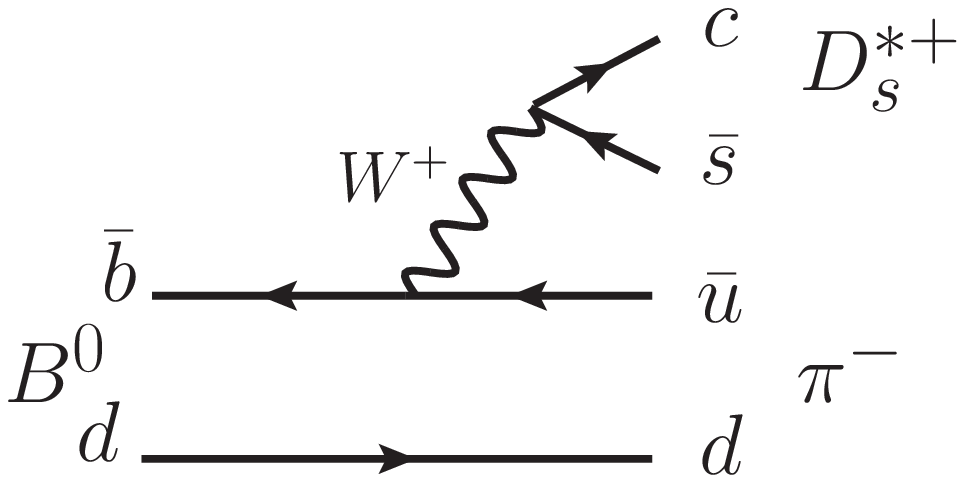}
\includegraphics[width=0.25\textwidth]{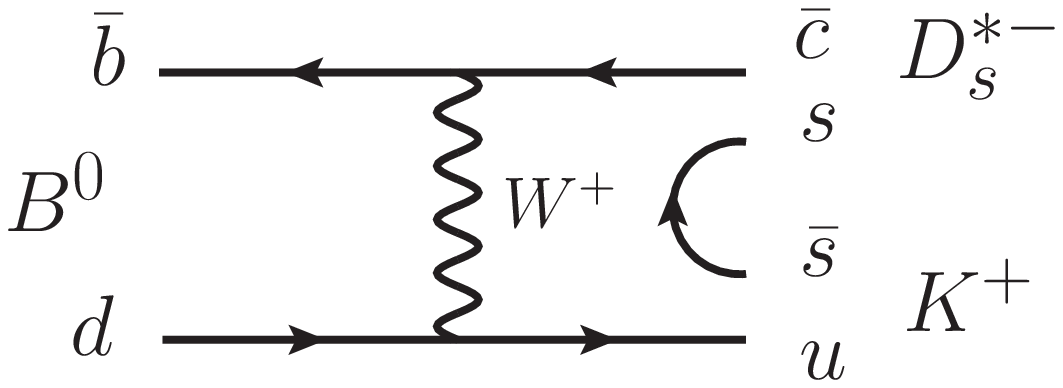}\\
\textrm{\hskip -1cm (c) \hskip 4cm (f)}
\caption{\label{decays}Feynman diagrams for (a) Cabibbo-favored decay
$B^0\to D^{*-}\pi^+$, (b) doubly Cabibbo-suppressed decay $B^0\to
D^{*+}\pi^-$, (c) SU(3) flavor symmetric $B^0\to D_s^{*+}\pi^-$; color
suppressed $W$-exchange contributions (d) to $B^0\to D^{*-}\pi^+$, (e)
to $B^0\to D^{*+}\pi^-$ and (f) to the decay $B^0\to D_s^{*-}K^+$.}
\end{figure}

 In the standard model (SM), $CP$ violation arises naturally when the
 Cabibbo-Kobayashi-Maskawa (CKM) mixing matrix is introduced into the
 weak-interaction Lagrangian~\cite{CKM}. A precise measurement of the
  CKM parameters is crucial for understanding $CP$ violation in the
   SM. In particular, the time-dependent $CP$ analysis of the $B^0
  (\bar{B}^0) \to D^{*\mp}\pi^{\pm}$ system provides a theoretically
    clean measurement of the product $R_{D^{*}\pi} \sin(2\phi_1 +
 \phi_3)$~\cite{sachs}, where $\phi_1$ and $\phi_3 $ are interior
 angles of the unitarity triangle and $R_{D^{*}\pi}$ is the ratio of
the magnitudes of the doubly Cabibbo-suppressed decay (DCSD) amplitude
 (Fig.~\ref{decays}(b)) to the Cabibbo-favored decay (CFD) amplitude
 (Fig.~\ref{decays}(a)). Measuring the DCSD amplitude is not possible
    with the current data, due to the overwhelming background from
 $\bar{B}^0 \to D^{*+}\pi^-$ and hence, it is not possible to extract
    the angle $\phi_3$ from this study alone unless an independent
measurement of $R_{D^{*}\pi}$ is provided externally. 
The mode $B^+ \to D^{*+}\pi^0$ may be used to estimate the size 
of DCSD, since $B^0 \to D^{*+}\pi^-$ and $B^+ \to D^{*+}\pi^0$ are 
related by isospin symmetry~\cite{sachs}. However, the 
$B^+ \to D^{*+}\pi^0$ branching fraction is small and so far 
only an upper limit has been obtained~\cite{iwabuchi}. Unlike the $B^0
\to D^{*\mp}\pi^\pm$ system, $B^0 \to D_s^{*+} \pi^-$, which is
predominantly a spectator process with a $b \to u$ transition
(Fig.~\ref{decays}(c)), does not have contributions from $\bar{B}^0$
decays to the same final state and can provide clean experimental
access to $R_{D^{*}\pi}$. Assuming SU(3) flavor symmetry between
$D^{*}$ and $D_s^*$, $R_{D^*\pi}$ is given by
\begin{equation}\label{eqn:r-d*pi}
R_{D^{*}\pi} = \tan \theta_C
\left(\frac{f_{D^{*}}}{f_{D_s^{*}}}\right) \sqrt{
\frac{\mathcal{B}(B^0 \to D_s^{*+}\pi^-)}{\mathcal{B}(B^0 \to
D^{*-}\pi^+)} },
\end{equation}
where $\theta_C$ is the Cabibbo angle, $f_{D^*}$ and $f_{D_s^*}$ are
the meson form factors, and the \BR's stand for the corresponding
branching fractions. The $B^0 \to D_s^{*+}\pi^-$ process, in addition,
does not have a penguin loop contribution and hence can in
principle be used to determine $|V_{ub}|$~\cite{kim}.

In contrast to the $B^0 \to D^{*\mp}\pi^{\pm}$ decays shown in
Figs.~\ref{decays}(d) and (e), the $B^0 \to D_s^{*+}\pi^-$ decay does
not have a contribution from the $W$-exchange amplitude, as the
quark-antiquark pair with the same flavor, required for such a
diagram, is absent from the final state. We assume the $W$-exchange
contributions in $B^0\to D^{*\mp}\pi^\pm$ to be negligible, in making
the correspondence between $D^{*+}\pi^-$ and $D_s^{*+}\pi^-$ in the
$R_{D^*\pi}$ calculation. The size of the $W$-exchange diagram can be
estimated from the $B^0 \to D_s^{*-} K^+$ decay, which proceeds only
via $W$-exchange (Fig.~\ref{decays}(f)). The $B^0 \to D_s^{*-} K^+$
branching fraction was expected to be enhanced due to rescattering
effects~\cite{block}.  However, a recent theoretical study based on
measurements of related processes indicates the absence of such an
enhancement~\cite{gron}.

While $B^0 \to D_s^+\pi^-$ and $B^0 \to D_s^-K^+$ decays have been
observed previously by Belle~\cite{krak} and BaBar~\cite{babar1}, the
observations of the modes $B^0 \to D_s^{*+}\pi^-$ and $B^0 \to
D_s^{*-}K^+$ have been reported by BaBar~\cite{babar2, babar3}, which
measured $\mathcal{B}(B^0 \to D_s^{*+}\pi^-) = (2.6^{+0.5}_{-0.4}\pm
0.3) \times 10^{-5}$ and $\mathcal{B}(B^0 \to D_s^{*-}K^+) = (2.4 \pm
0.4 \pm 0.2) \times 10^{-5}$. In this paper, we report an improved
measurement of the branching fractions for the decays \dsstPi~ and
\dsstK~\cite{CC} with a data sample consisting of $657 \times 10^6$
$B\bar{B}$ pairs, collected with the Belle detector at the KEKB
asymmetric-energy $e^+e^-$ collider~\cite{KEKB}.

The Belle detector is a large-solid-angle magnetic spectrometer that
consists of a silicon vertex detector, a 50-layer central drift
chamber (CDC), an array of aerogel threshold Cherenkov counters (ACC),
a barrel-like arrangement of time-of-flight scintillation counters
(TOF), and an electromagnetic calorimeter comprised of CsI(Tl)
crystals located inside a superconducting solenoid coil that provides
a 1.5 T magnetic field. An iron flux-return yoke located outside the
solenoid is instrumented to detect $K^0_L$ mesons and to identify
muons. The detector is described in detail elsewhere~\cite{Belle}. Two
different inner detector configurations were used.  For the first
sample of $152 \times 10^6$ $B\bar{B}$ pairs, a $2.0$ cm radius
beam-pipe and a 3-layer silicon vertex detector were used; for the
latter $505 \times 10^6$ $B\bar{B}$ pairs, a $1.5$ cm radius beam-pipe
with a 4-layer silicon vertex detector and a small-cell inner drift
chamber were used~ \cite{svd2}.

The signal is reconstructed in three \dsub{+}~modes: $\phi\pi^+$ with
$\phi \to K^+K^-$, $\bar{K}^*(892)^0K^+$ with $\bar{K}^*(892)^0 \to
K^-\pi^+$, and $K_S^0K^+$ with $K_S^0 \to \pi^+\pi^-$.  Charged tracks
are selected with requirements based on the impact parameter relative
to the interaction point (IP). The deviations from the IP are required
to be within $\pm 4$ cm along the $z$-axis (the direction opposite to
the positron beam) and within 0.2 cm in the $x-y$ plane.  We also
require the transverse momentum of the tracks to be greater than 0.1
GeV/$c$ in order to reduce low momentum combinatorial background.

For charged particle identification (PID), we combine the information
from the specific ionization ($dE/dx$) in the CDC with measurements
from the TOF and ACC. At large momenta ($p > 2.5$ GeV/$c$) only the
ACC measurement and $dE/dx$ are used. We assign likelihood values
${\cal L}_K$ (${\cal L}_\pi$) for the kaon (pion) hypothesis to each
charged track. Tracks are identified based on the ratio ${\cal
R}_{K/\pi} = {\cal L}_K/ ({\cal L}_K + {\cal L}_\pi)$, which peaks at
one for real kaons and at zero for real pions.  For the prompt 
kaon (pion) track, we require ${\cal R}_{K/\pi} > 0.6 ( < 0.6)$, for
which the identification efficiency is 85\% (92\%) with a pion (kaon)
fake-rate of 8\% (15\%). Due to the low background level for modes
with a $\phi$ meson, less restrictive PID cuts of ${\cal R}_{K/\pi} >
0.2$ are applied to the $\phi$ daughter tracks.

The $\phi$ (\kstar) mesons are required to have an invariant mass
within $\pm 14$ MeV/$c^2$ ($\pm75$~MeV/$c^2$) of the nominal $\phi$
(\kstar) mass~\cite{PDG}. We reconstruct \kshort~candidates
from $\pi^+\pi^-$ pairs, requiring the invariant mass to be within
$\pm 10$ MeV/$c^2$ ($\sim \pm 3\sigma$) of the nominal \kshort~mass.
The \kshort~candidate is further required to pass a momentum-dependent
selection criteria based on its vertex topology, the flight length in
the $r-\phi$ plane, and the daughter $\pi^\pm$ momentum
distribution~\cite{k.f.chen}. The \dsub{+}~candidate mass window for
$\phi\pi^+$, $\bar{K}^*(892)^0K^+$, and $K_S^0K^+$ modes is $\pm 13$
MeV/$c^2$, $\pm 15$~MeV/$c^2$, and $\pm 17$ MeV/$c^2$, respectively;
these ranges correspond to approximately 3$\sigma$ in resolution
around the $D_s^+$ mass. To reduce combinatorial background, 
we use a more stringent PID requirement for the kaon
accompanying the $\bar{K}^*(892)^0$, ${\cal R}_{K/\pi} > 0.8$.  The
\dsub{+}~candidate is constrained kinematically to have a mass
equal to the nominal value~\cite{PDG}.

The \dsst{+}~mesons are reconstructed by combining the
\dsub{+}~candidates with a photon. The photons are reconstructed from
energy depositions in the ECL and are required to have energies
greater than $60$ MeV ($100$ MeV) in the barrel (endcap) region
covering the polar angle $32^\circ < \theta < 128^\circ$ ($17^\circ <
\theta < 32^\circ$ (forward endcap) and $128^\circ < \theta <
150^\circ$ (backward endcap)). The \dsst{+} ~candidate is required to
have $\Delta M = M_{D_s^{+}\gamma} - M_{D_s^+}$ between $128 \textrm{
MeV/c}^2$ and $162 \textrm{ MeV/c}^2$, where $M_{D_s^{+}\gamma}$ and
$M_{D_s^+}$ are the invariant masses of the $D_s^+\gamma$ system and
the $D_s^+$ candidate, respectively.  To reduce the combinatorial
background due to low energy photons, we require that
$\cos\theta_{D_s^{*+}} > -0.6$ ($-0.7$) for
\dsstPi~(\dsstK), where $\theta_{D_s^{*+}}$ is defined as the angle
between the flight direction of the photon and the direction opposite
to the $B^0$ flight in the \dsst{+}~rest frame.  We then perform a
mass-constrained fit to the \dsst{+}~candidate.  This improves the
momentum resolution by 25\%.

The $B^0$ candidates, reconstructed by combining a \dsst{+} candidate
with an oppositely charged pion/kaon track, are identified by the
energy difference, $\Delta E = \sum_i E_i - E_{\rm beam}$ and the
beam-energy constrained mass, $M_{\rm bc} = \sqrt{E_{\rm beam}^2 -
(\sum_i \vec{p_i})^2}$, where $E_{\rm beam}$ is the beam energy in the
$\Upsilon(4S)$ center-of-mass (CM) frame and $\vec{p_i}$ and $E_i$ are
the momentum and energy of the $i$th daughter of the $B^0$ in the CM
frame. We retain $B^0$ candidates with $\Delta E$ within $\pm 0.2$ GeV
and $M_{\rm bc}$ between $5.2$~GeV/$c^2$ and $5.3$ GeV/$c^2$ for
further analysis.

The dominant background comes from the $e^+e^- \to q\bar{q}$ ($q=u,d,s
\textrm{ and } c$ quarks) continuum process. To suppress this
background, we use the event topology in the CM frame to distinguish
more spherical $B\bar{B}$ events from the jet-like continuum events. A
likelihood function ${\cal R} = {\cal L}_{\rm sig} / ({\cal L}_{\rm
sig}+ {\cal L}_ {\rm bkg})$ is prepared by combining a Fisher
discriminant, based on a set of modified Fox-Wolfram
moments~\cite{SFW,KSFW} with $\cos\theta_B$, where $\theta_B$ is
the polar angle of the $B^0$ meson flight direction in the CM
frame. The angle $\theta_B$ follows a $\sin^2\theta_B$ distribution
 for $B\bar{B}$ events, while the continuum
distribution is flat. The selection criteria for ${\cal R}$ are
determined by maximizing a figure-of-merit, $S/\sqrt{S+B}$, where $S$
and $B$ are the number of signal and background events determined from
large Monte Carlo (MC) samples~\cite{genSim}, with statistics
corresponding to about 100 (5) times data for signal
(background) MC.  The signal yield, $S$ is obtained assuming the
latest branching fraction measurements~\cite{PDG}.  In the case of
\dsstPi~(\dsstK), we require ${\cal R}$ to be greater than 0.45 (0.45)
for the \phiMode, 0.50 (0.60) for the \kstarMode, and 0.40 (0.40) for
the \kshortMode. For the \phiMode, this requirement 
suppresses 80\% of the continuum background, while retaining 85\% 
of the signal.

About 15$\%$ of events have more than one $B^0$ signal candidate. For
these events we choose the candidate with the $M_{\rm bc}$ value
closest to the nominal $B^0$ mass. This procedure selects the correct
$B^0$ candidate in about 92\% of the cases. Only events with $M_{\rm
bc}$ between 5.27 GeV/$c^2$ and 5.29~GeV/$c^2$ are considered for
further analysis, while the signal is extracted by performing a fit to
the $\Delta E$ distribution. We define the fit region to be $|\Delta
E| < 0.2$ GeV.

A MC sample of $B\bar{B}$ events is used to determine possible
backgrounds that can enter the $\Delta E$ fit region. In both signal
modes ($B^0 \to D_s^*h$, where $h$ is a charged $K$ or $\pi$), about
45\% of the background comes from decays involving a $D^+ \to
K^-\pi^+\pi^+$ or a $D^+ \to K_S^0\pi^+$ sub-decay, which form a fake
\dsub{+} when the $\pi^+$ from the $D^+$ is misidentified as a
$K^+$. However, these events do not peak and are distributed over the
entire fit region due to the addition of a random photon.

On the other hand, some rare $B$ decays to final states that contain a
correctly reconstructed $D_s^{(*)+}$ produce non-negligible peaking
structures in the $\Delta E$ fit region: $B^0 \to D_s^+\pi^-$ ($B^0
\to D_s^-K^+$) events populate the region around 150 MeV due to the
addition of an extra photon, $B^0 \to D_s^{*+}\rho^-$ ($B^+ \to
D_s^{*-}K^+\pi^+$) events populate the region around $-150$ MeV, since
a $\pi^0$ ($\pi^+$) is not reconstructed, while the events 
from $B^0 \to D_s^{+}\rho^-$ ($B^+ \to D_s^{-}K^+\pi^+$)
are distributed around $-50$ MeV, with the extra photon compensating
the lost pion. These backgrounds are represented by PDFs with fixed
yields. MC samples are used to determine the PDF parameters as well as
efficiencies. The yields are calculated assuming the most recent known
values for their branching fractions~\cite{PDG,babar3}. Apart from the
backgrounds discussed above, the two $B^0$ signal modes cross-feed
each other. We use signal MCs to determine the PDFs for the
cross-feeds. $\Delta M$ sidebands in the data are used to verify the
consistency of the MC background predictions with the data.

The signal PDF is the sum of a Crystal Ball line-shape~\cite{cbshape}
and a broad Gaussian and is parametrized using signal MC samples
generated in each \dsub{+}~mode. The signal as well as the peaking
background PDFs are subsequently corrected for possible differences in
the parameter vaues between MC and real data, using a $B^0 \to
D_s^{*+}D^-$ data control sample. The combinatorial backgrounds in
each mode are accounted for by adding linear functions.

\begin{figure}[h!h!]
\begin{center}
\includegraphics[width=0.5\textwidth]{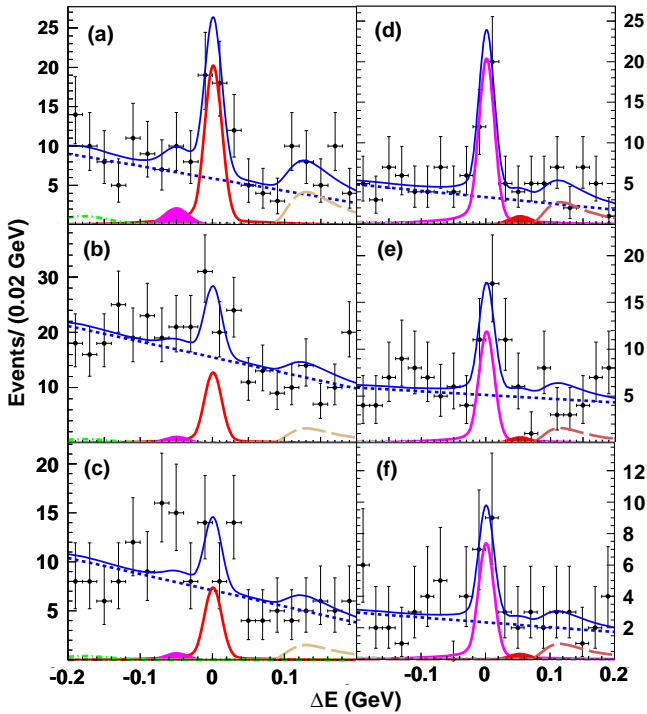}
\caption{\label{fig:result} The simultaneous fit in the $B^0 \to
 D_s^{*+}\pi^-$ ((a) $\phi\pi$, (b) $\bar{K}^{*0}K$ and (c) $K_S^0K$
 mode) and the $B^0 \to D_s^{*-}K^+$ ((d)-(f)) signal modes. Signal
 peaks are shown by the solid curves while the solid-filled curves
 represent the cross-feed contributions from the other $B^0$ signal
 modes. The long-dashed curves correspond to the contribution from the
 $B \to D_s\pi$ ($B \to D_sK$) and the dot-dashed curves to that from
 $B^0 \to D_s^{(*)+}\rho^-$ ($B^+ \to D_s^{(*)-}K^+\pi^+$). The dotted
 curves correspond to the combinatorial background.}
\end{center}
\end{figure}

We determine the branching fractions from a simultaneous unbinned
extended maximum likelihood fit to the $\Delta E$ distributions of the
three \dsub{+} decay modes for each signal mode.  To account for the
cross-feeds between the signal modes due to the misidentification of
the prompt track, the two $B$ signal modes are fitted simultaneously,
with the \dsstK~signal yield in the correctly reconstructed sample
determining the normalization of the cross-feed in the \dsstPi fit
region, and vice versa.  The fit has $14$ free parameters: the
branching fractions of the signal modes ($2$), and the yields and
slopes of the first-order polynomials representing the combinatorial
background in each of the three
\dsub{+}~modes~($12$). Figure~\ref{fig:result} shows results of the
simultaneous fit for the three \dsub{+} modes in both signal modes.~
We summarize the results of the fits in Table~\ref{tab:result}. The
significance is defined as $\sqrt{-2\ln ({\cal L}_0 / {\cal L}_{\rm
max})}$, where ${\cal L}_{\rm max}$ (${\cal L}_0$) are the likelihoods
for the best fit and with the signal branching fraction fixed to zero.

\begin{table*}[t!]
\caption{\label{tab:result}Efficiency ($\epsilon$), yield $(N_{\rm
sig})$, branching fraction $({\cal B})$ and statistical significance
not including systematic errors ($\mathcal{S}$) from the fits to the
data obtained individually in the three $D_s^+$ modes as well as from
the simultaneous fit. The second error on the \BR's is due to the
uncertainties in $D_s^+$ decay branching fractions. The individual fit
results are consistent with each other and also with the simultaneous
fit. Note that the efficiencies in the \kshortMode~include the $K_S^0
\to \pi^+\pi^-$ decay branching fraction.}

\begin{tabular}{llp{0.5cm}cp{0.5cm}cp{0.5cm}cp{0.5cm}c}

\hline\hline

$B^0$ mode 
& $D_s^+$ mode
&
& $\epsilon\;(\%)$
&
& $N_{\rm sig}$ 
&
& \multicolumn{1}{c}{${\cal B}(10^{-5})$} 
&
& $\mathcal{S}\; (\sigma)$\\

\hline

\multirow{4}{*}{\dsstPi} 
& $\phi(K^+K^-)\pi^+$
&
& $15.2$
&
& $32\pm 8$
&
& $1.58 \pm 0.40 \pm 0.24$ 
& 
& $3.2$ \\

& $\bar{K}^*(892)^0(K^-\pi^+)K^+$
&
& $7.9$
& 
& $29\pm 10$
&
& $2.30 \pm 0.76 \pm 0.35$
& 
& $2.6$ \\

& $K_S^0K^+$
&
& $8.0$
& 
& $13 \pm 7$
&
& $1.78 \pm 0.92 \pm 0.11$
& 
& $2.2$ \\

& Simultaneous
&
& -
& 
& -
&
& $1.75 \pm 0.34 \pm 0.11$ 
& 
& $6.6$ \\

\hline

\multirow{4}{*}{\dsstK} 
& $\phi(K^+K^-)\pi^+$ 
&
& $13.4$
& 
& $33\pm 8$
&
& $1.81 \pm 0.41 \pm 0.27$ 
& 
& $3.2$ \\

& $\bar{K}^*(892)^0(K^-\pi^+)K^+$
&
& $6.4$
& 
& $23 \pm 7$
&
& $2.22 \pm 0.66 \pm 0.34$
& 
& $2.8$ \\

& $K_S^0K^+$
&
& $6.9$
& 
& $14 \pm 5$
&
& $2.14 \pm 0.80 \pm 0.13$
& 
& $3.1$ \\

& Simultaneous
&
& -
& 
& -
&
& $2.02 \pm 0.33 \pm 0.13$ 
& 
& $8.6$ \\

\hline\hline

\end{tabular}
\end{table*}

%\section{Systematic Uncertainty}
The major source of systematic uncertainty in the branching fraction
measurement of \dsstPi (\dsstK) is the uncertainty in the branching
fractions of the \dsub{+}~decays, which amount to 5.9\% (6.2\%).  The
uncertainties in the branching fractions of the peaking background
modes contribute an additional error of 1.5\% (1.9\%). The systematic
uncertainty in the tracking efficiency is estimated to be about 1.0\%
per track. The uncertainty in the PID efficiency is about 2.4\%
(2.1\%). Photon detection efficiency has an uncertainty of 7.0\%,
while \kshort~ detection efficiency adds 1.1\% uncertainty in the
result. The efficiency of the ${\cal R}$ requirement, 
used to suppress the continuum background, introduces an uncertainty 
of 0.6\% (0.5\%) in the branching fraction. The limited size of the MC samples
used to determine efficiencies and cross-feed fractions introduces an
error of 1.4\% (1.6\%). The uncertainty in the determination of the
signal PDF shape is about 3.4\% (1.5\%).  Estimation of possible bias
in the fit results in another 0.9\% (0.3\%) uncertainty.

Table~\ref{tab:syst} summarizes the systematic uncertainties
involved. The overall systematic error is obtained by adding the above
contributions in quadrature.

\begin{table}[h!]
  \caption{\label{tab:syst} Contributions to the
  systematic uncertainty}

  \begin{tabular}{lcc}
    \hline \hline Source & \multicolumn{2}{c}{Contribution(\%)} \\
    
    & $D_s^{*+}\pi^-$ & $D_s^{*+}K^-$ \\
    
    \hline
    
    \multicolumn{3}{l}{\bf $\mathbf D_s^+$ branching fraction
    uncertainties} \\
    
    signal & 5.9 & 6.2 \\ 

    peaking background & 1.5 & 1.9 \\
    
    \multicolumn{3}{c}{\rule{0.4\textwidth}{0.3mm}} \\
    
		{\bf Total (${\cal B}$)} & $\mathbf{6.1}$ & 
		$\mathbf{6.5}$ \\
		
		\hline
		
		Tracking efficiency & 4.0 & 4.0 \\
		Photon detection efficiency & 7.0 & 7.0 \\
		Particle identification efficiency & 2.4 & 2.1 \\
		$K_S^0$ efficiency &  1.1 & 1.1 \\
		${\cal LR}$ & 0.6 & 0.5 \\
		$N_{B\bar{B}}$ & 1.4 & 1.4 \\
		MC statistics & 1.4 & 1.6 \\
		PDF shape & 3.4 & 1.5 \\
		Fit bias & 0.9 & 0.3 \\
		
		\multicolumn{3}{c}{\rule{0.4\textwidth}{0.3mm}} \\
		
			    {\bf Total (other)} & $\mathbf{9.4}$ &
			    $\mathbf{8.8}$ \\

			    \hline\hline
			    
  \end{tabular}
\end{table}

We obtain \resDsstPi~ and \resDsstK~ with significances of $6.1\sigma$
and $8.0\sigma$, respectively, where the systematic uncertainties on
the signal yield as well as the statistical uncertainties are included
in the significance evaluation. Though consistent with the previous
measurements~\cite{babar3}, we observe slightly lower branching
fractions. Using the observed value for the $B^0 \to D_s^{*+}\pi^-$
branching fraction, the latest values for ${\cal B}(B^0 \to
D^{*-}\pi^+) = (2.76 \pm 0.13) \times 10^{-3}$, $\tan\theta_C = 0.2314
\pm 0.0021$ ~\cite{PDG}, and the theoretical estimate of the ratio
$f_{D_s^{+}}/f_{D^{+}} = ~(1.164 \pm 0.006 \textrm{ (stat) } \pm 0.020
\textrm{ (syst)})$~\cite{follana}, we obtain,

\begin{equation}\nonumber
R_{D^*\pi} = (1.58 \pm 0.15 (\textrm{stat}) \pm 0.10 (\textrm{syst})
\pm 0.03 (\textrm{th}))\%,
\end{equation}
where the first error is statistical, the second corresponds to the
experimental systematic uncertainty and the third accounts for the
theoretical uncertainty in the $f_{D_s^{+}}/f_{D^{+}}$ estimation. We
have assumed that the ratio $f_{D_s}/f_D$ is equal to the ratio of 
vector meson decay constants, $f_{D_s^*}/f_{D^*}$.  The
quenched QCD approximation~\cite{becirevic} as well as the heavy quark
effective theory predictions~\cite{neubert} point toward an
uncertainty of about 1\% due to this assumption, which is included in
our estimation of $R_{D^*\pi}$. The value we obtain for $R_{D^*\pi}$,
though consistent with the theoretical expectation of $2\%$, is
slightly smaller than the previous estimate~\cite{babar3}.

The observed value for the \dsstK~ branching fraction is two orders 
of magnitude lower than that for the
Cabibbo-favored decay $B^0 \to D^{*-}\pi^+$. This can be understood
purely in terms of the exchange amplitude and there is no evidence for
enhancement due to rescattering effects, which would lead to
comparable amplitudes for the two processes~\cite{block}. From this
same comparison, we find no evidence for large $W$-exchange
contributions to $B^0 \to D^{*\mp}\pi^\pm$; such contributions are
assumed to be small, in the determination of $R_{D^*\pi}$
(Eq.~\ref{eqn:r-d*pi}).

In conclusion, we report the most precise measurement of the
\dsstPi~and \dsstK decay branching fractions. This improves the
precision with which the parameter $R_{D^*\pi}$ can be estimated, and
thus the prospect of determining $\phi_3$ from measurements of $CP$
violating effects in the $D^{*\pm}\pi^\mp$ system.

We thank the KEKB group for the excellent operation of the
accelerator, the KEK cryogenics group for the efficient operation of
the solenoid, and the KEK computer group and the National Institute of
Informatics for valuable computing and SINET3 network support.  We
acknowledge support from the Ministry of Education, Culture, Sports,
Science, and Technology (MEXT) of Japan, the Japan Society for the
Promotion of Science (JSPS), and the Tau-Lepton Physics Research
Center of Nagoya University; the Australian Research Council and the
Australian Department of Industry, Innovation, Science and Research;
the National Natural Science Foundation of China under contract
No.~10575109, 10775142, 10875115 and 10825524; the Department of
Science and Technology of India; the BK21 and WCU program of the
Ministry Education Science and Technology, the CHEP SRC program and
Basic Research program (grant No.  R01-2008-000-10477-0) of the Korea
Science and Engineering Foundation, Korea Research Foundation
(KRF-2008-313-C00177), and the Korea Institute of Science and
Technology Information; the Polish Ministry of Science and Higher
Education; the Ministry of Education and Science of the Russian
Federation and the Russian Federal Agency for Atomic Energy; the
Slovenian Research Agency; the Swiss National Science Foundation; the
National Science Council and the Ministry of Education of Taiwan; and
the U.S.\ Department of Energy.  This work is supported by a
Grant-in-Aid from MEXT for Science Research in a Priority Area (``New
Development of Flavor Physics''), and from JSPS for Creative
Scientific Research (``Evolution of Tau-lepton Physics''). Author
N.J.J. thanks Prof. Kazuo Abe of IPMU for illuminating discussions and
guidance during the initial development of this work.


\begin{thebibliography}{99}
\bibitem{CC} Inclusion of the charge-conjugate states is implicit
throughout this work unless otherwise stated.
\end{thebibliography}
\end{document}